\documentclass[12pt,preprint]{aastex}
\shorttitle{}
\shortauthors{Nesvorn\'y et al.}
\begin{document}
\title{The orbital distribution of trans-Neptunian objects \\ beyond 50~au}
\author{David Nesvorn\'y$^1$, David Vokrouhlick\'y$^2$, Fernando Roig$^3$}
\affil{(1) Department of Space Studies, Southwest Research Institute, 1050 Walnut St., \\Suite 300, 
Boulder, CO 80302, USA} 
\affil{(2) Institute of Astronomy, Charles University, V Hole\v{s}ovi\v{c}k\'ach 2, CZ--18000 Prague 8,
Czech Republic} 
\affil{(3) Observat\'orio Nacional, Rua Gal. Jose Cristino 77, Rio de Janeiro, RJ 20921-400, Brazil}

\begin{abstract}
The dynamical structure of the Kuiper belt beyond 50 au is not well understood. Here we report results 
of a numerical model with long-range, slow and grainy migration of Neptune. The model implies that bodies scattered 
outward by Neptune to semimajor axes $a>50$ au often evolve into resonances which subsequently act to 
raise the perihelion distances of orbits to $q>40$ au. The implication of the model is that the orbits
with $50<a<100$ au and $q>40$ au should cluster near (but not in) the resonances with Neptune (3:1 at $a=62.6$ 
au, 4:1 at $a=75.9$ au, 5:1 at $a=88.0$ au, etc.). The recent detection of several distant Kuiper Belt 
Objects (KBOs) near resonances is consistent with this prediction, but it is not yet clear whether the
orbits are really non-resonant as our model predicts. We estimate from the model that there should presently 
be $\sim$1600-2400 bodies at the 3:1 resonance and $\sim$1000-1400 bodies at the 4:1 resonance 
(for $q>40$ au and diameters $D>100$ km). These results favorably compare with the population census
of distant KBOs inferred from existing observations. 
\end{abstract}
\section{Introduction}
In our previous work, we developed a numerical model of Neptune's migration into an outer planetesimal 
disk (Nesvorn\'y 2015a,b; Nesvorn\'y \& Vokrouhlick\'y 2016; hereafter NV16). By comparing the model results 
with the observed distribution of Kuiper belt orbits with $a<50$ au (e.g., Petit et al. 2011), we inferred 
that Neptune's migration must have been long-range, slow and grainy. Here we use the same model to discuss 
the orbital structure of the Kuiper belt beyond 50 au. We find that objects scattered by Neptune to $a>50$~au 
are often trapped into mean motion resonances with Neptune which act to raise the perihelion distances to $q>40$ 
au, and detach the orbits from Neptune. The objects are subsequently released from resonances as Neptune 
migrates toward its present orbit. The orbital structure of the detached disk with $a>50$ au and $q>40$ au is thus 
expected to be clustered near Neptune's resonances. Similar results were recently reported in an independent 
work (Kaib \& Sheppard 2016). Section 2 briefly reviews the numerical method. The results are presented and 
compared with observations in Section 3. Our conclusions are given in Section~4.  
\section{Method}
{\it Integration Method.} Our numerical integrations consist of tracking the orbits of four giant planets 
(Jupiter to Neptune) and a large number of particles representing the outer planetesimal disk. To set up an 
integration, Jupiter and Saturn are placed on their current orbits.
Uranus and Neptune are placed inside of their current orbits and are migrated outward. The initial 
semimajor axis $a_{\rm N,0}$, eccentricity $e_{\rm N,0}$, and inclination $i_{\rm N,0}$ define Neptune's orbit before 
the main stage of migration/instability. The {\tt swift\_rmvs4} code, part of the {\it Swift} $N$-body integration 
package (Levison \& Duncan 1994), is used to follow the orbital evolution of all bodies. 

The code was modified to include artificial forces that mimic the radial migration and damping of planetary orbits. 
These forces are parametrized by the exponential e-folding timescales, $\tau_a$, $\tau_e$ and $\tau_i$, where $\tau_a$ 
controls the radial migration rate, and $\tau_e$ and $\tau_i$ control the damping rates of $e$ and $i$ (NV16). 
We set $\tau_a=\tau_e=\tau_i$ because such roughly comparable timescales were suggested by previous work.
The numerical integration is divided into two stages with migration/damping timescales $\tau_1$ and $\tau_2$ (NV16).  
The first migration stage is stopped when Neptune reaches $a_{\rm N,1}\simeq27.7$ au. 
Then, to approximate the effect of planetary encounters during dynamical instability, we apply a discontinuous change 
of Neptune's semimajor axis and eccentricity, $\Delta a_{\rm N}$ and $\Delta e_{\rm N}$. Motivated by previous results
(Nesvorn\'y \& Morbidelli 2012, hereafter NM12), we set $\Delta a_{\rm N}=0.5$~au and $\Delta e_{\rm N}=0.1$. 

The second migration stage starts with Neptune having the semimajor axis $a_{\rm N,2}=a_{\rm N,1}+\Delta a_{\rm N}$. 
We use the {\tt swift\_rmvs4} code, and migrate the semimajor axis (and damp the eccentricity) on an e-folding 
timescale $\tau_2$. The migration amplitude was adjusted such that the planetary orbits obtained at the end 
of the simulations were nearly identical to the real orbits. This guarantees that the mean motion and secular resonances 
reach their present positions. 

We found from NM12 that the orbital behavior of Neptune during the first and second migration 
stages can be approximated by $\tau_1\simeq10$ Myr and $\tau_2\simeq30$~Myr for a disk mass $M_{\rm disk}=20$ $M_{\rm Earth}$, 
and $\tau_1\simeq20$~Myr and $\tau_2\simeq50$ Myr for $M_{\rm disk}=15$~$M_{\rm Earth}$. The real migration slows down,
relative to a simple exponential, at late stages. We therefore use $\tau_1=10$-30 Myr and $\tau_2=30$-100 Myr. All 
migration simulations were run to 0.5~Gyr. They were extended to 4.5~Gyr with the standard {\tt swift\_rmvs4} code 
(i.e., without migration/damping after 0.5~Gyr). 

{\it Migration graininess.}
We developed an approximate method to represent the jitter that Neptune's orbit experiences due to close
encounters with massive planetesimals. The method has the flexibility to use any smooth migration history of Neptune 
as an input, include any number of massive planetesimals in the original disk, and generate a new migration 
history where the random element of encounters with the massive planetesimals is included. This approach is useful, because 
we can easily control how grainy the migration is while preserving the global orbital evolution of planets from the 
smooth simulations. See NV16 for a detailed description of the method. Here we set the mass of massive planetesimals 
to be equal to that of Pluto. We motivate this choice by the fact that two Pluto-class objects are known in the Kuiper
belt today (Pluto and Eris). See NV16 for a discussion. 

{\it Planetesimal Disk.} 
The planetesimal disk is divided into two parts. The part from just outside Neptune's initial orbit to 
$r_{\rm edge}$ is assumed to represent the massive inner part of the disk (NM12). We use $r_{\rm edge}=28$-30 au, 
because our previous simulations in NM12 showed that the massive disk's edge must be at 28-30 au for Neptune to stop at $\simeq$30~au
(Gomes et al. 2004). The estimated mass of the planetesimal disk below 30~au is $M_{\rm disk}\simeq15$-20~$M_{\rm Earth}$ 
(NM12). The massive disk has a crucial importance here, because it is the main source of the resonant populations, Hot 
Classicals and Scattered Disk Objects (SDOs) (e.g., Levison et al. 2008). The planetesimal disk had a low 
mass extension reaching from 30 au to at least $\simeq$45 au. The disk extension is needed to explain why the 
Cold Classicals have several unique physical and orbital properties, but it does not substantially contribute 
to the SDOs, because of the small original mass of the extension. Here we therefore ignore the outer extension of 
the disk. 

Each of our simulations includes one million disk particles distributed from outside Neptune's initial orbit to 
$r_{\rm edge}$. The radial profile is set such that the disk surface density $\Sigma \propto 1/r$, where $r$ is 
the heliocentric distance. The initial eccentricities and initial inclinations of disk particles in 
our simulations are distributed according to the Rayleigh distribution (Nesvorn\'y 2015a).
The disk particles are assumed to be massless, such that their gravity does not interfere with the migration/damping 
routines. This means that the precession frequencies of planets are not affected by the disk in our simulations, while 
in reality they were (Batygin et al. 2011). 

{\it Effects of other planets.} 
The gravitational effects of the fifth giant planet (NM12) and planet 9 (Trujillo \& Sheppard 2014, Batygin \& Brown 2016) 
on the disk planetesimals are ignored. The fifth giant planet is short lived and not likely to cause major perturbations
of orbits in the Kuiper belt (although this may depend on how exactly planets evolve during the instability; e.g., Batygin 
et al. 2012). Given its presumably wide orbit, planet 9 does not affect orbits with $a<100$ AU, but may have a major influence 
on the structure of the scattered disk above 100 AU (e.g., Lawler et al. 2016). We therefore focus on the 50-100 au region 
in this work. 
\section{Results}
Here we report the results of two selected simulations from NV16. 
The first one (Case~1) corresponds to $\tau_1=30$ Myr, $\tau_2=100$ Myr, 
$\Delta a_{\rm N}=0.5$ au, and 4000 Pluto-mass objects in the original planetesimal disk. The second one (Case 2) has 
$\tau_1=10$ Myr, $\tau_2=30$ Myr, $\Delta a_{\rm N}=0.5$ au, and 1000 Pluto-mass objects. We used a larger number of Plutos 
in Case 1 than in Case 2, because there is some trade off between the migration graininess and speed. Both these simulations 
were shown to reproduce the correct architecture of the Kuiper belt below 50 au (NV16).

\begin{figure}
\epsscale{0.6}
\plotone{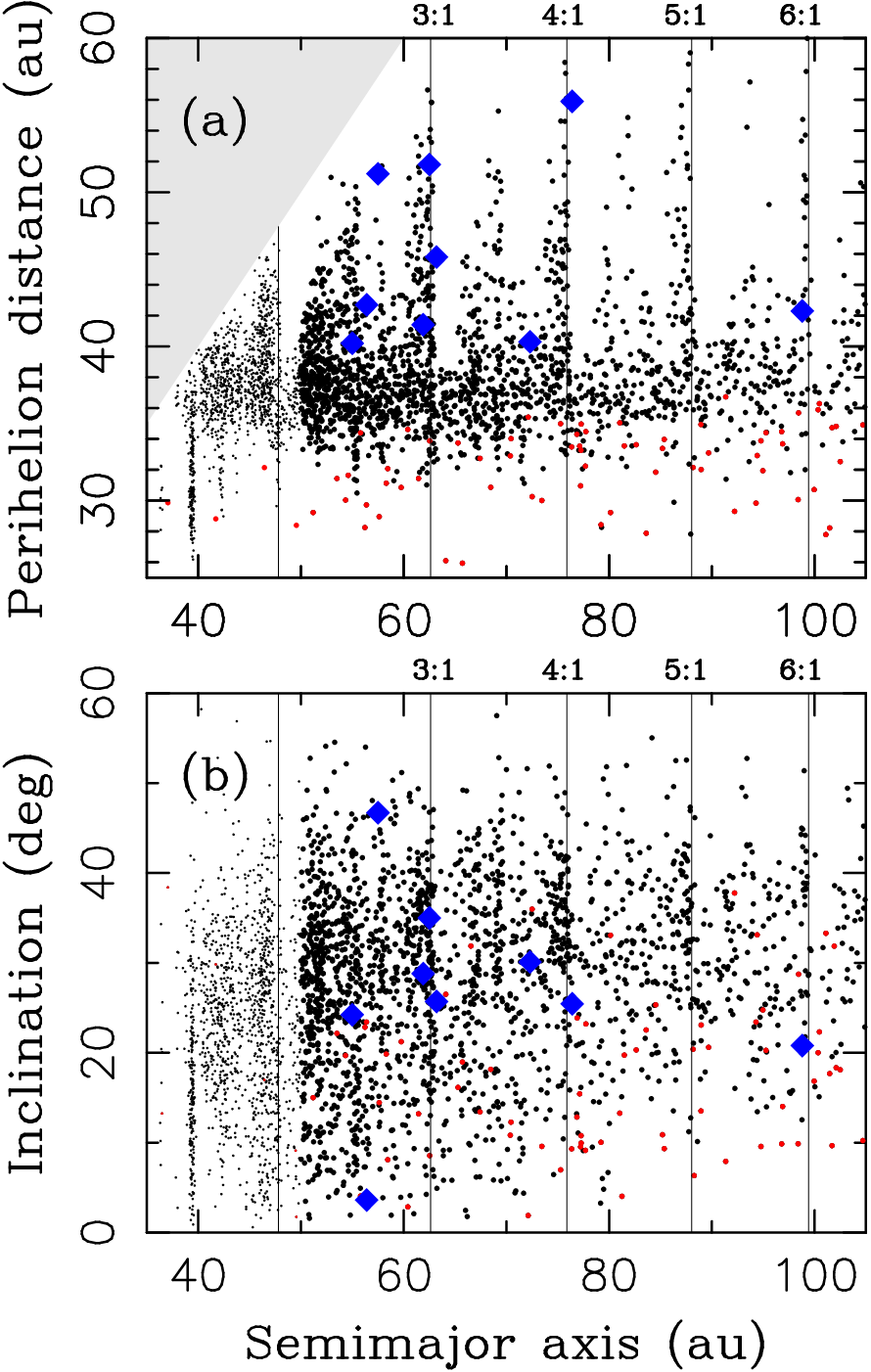}
\caption{The orbital distribution of bodies produced in our Case-1 simulation ($\tau_1=30$~Myr, $\tau_2=100$ Myr, 4000 Plutos).
The upper (lower) panel shows the perihelion distance (inclination). The orbits with $a>50$ au, which are the main focus here, 
are denoted by larger dots. The scattering orbits, defined as those whose semimajor axis changed more than 1.5~au in a 
10-Myr integration (Gladman et al. 2008), are denoted by red dots. Note the massive detached population (black dots). 
The detached objects with $q>40$ au are concentrated near resonances. The known KBOs with $q>40$ au, 
reported in Table 1 of Sheppard et al. (2016), are shown by blue diamonds. The orbital elements plotted here are 
barycentric.}
\label{case1}
\end{figure}

\begin{figure}
\epsscale{0.6}
\plotone{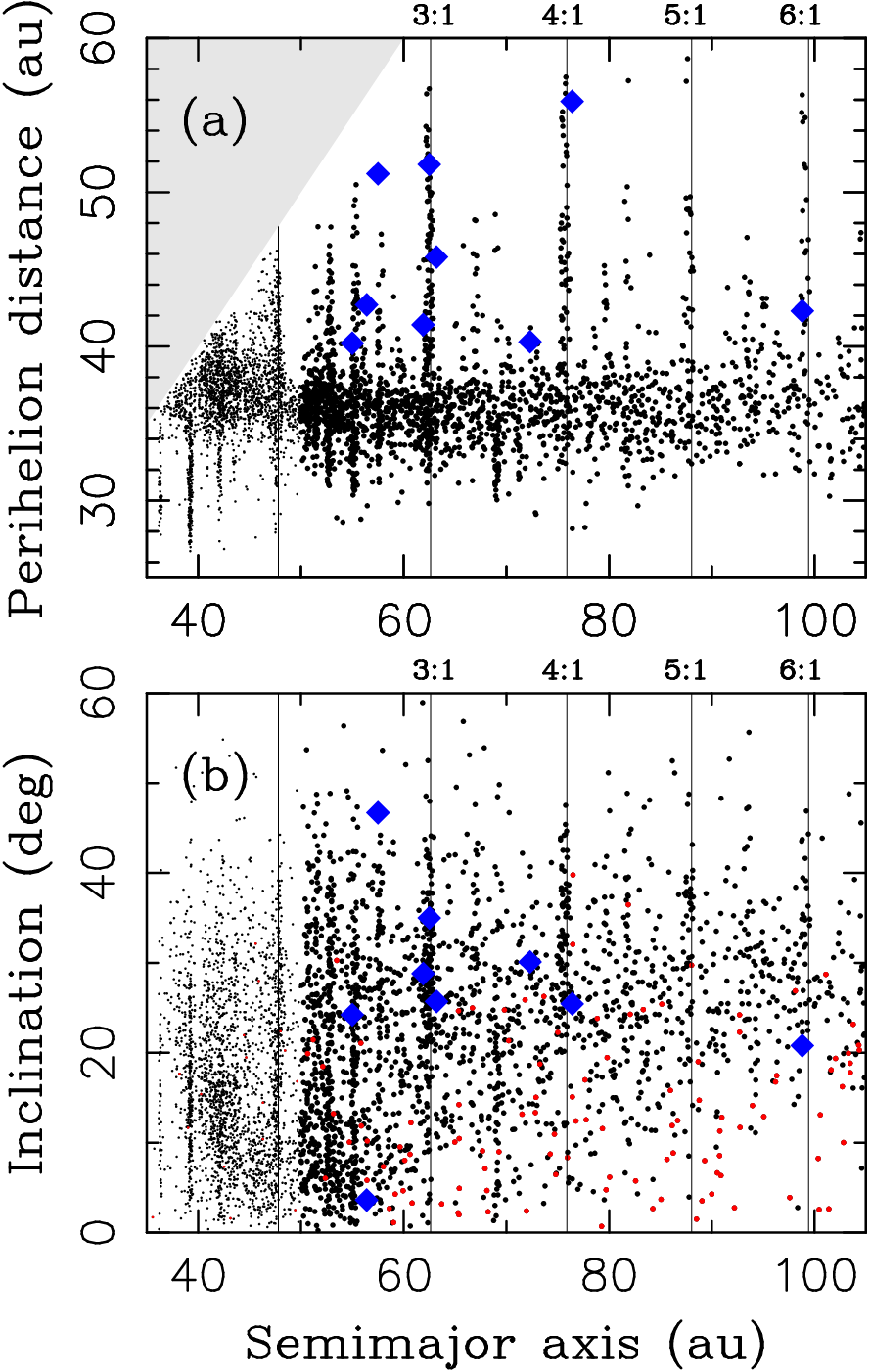}
\caption{The same as Figure \ref{case1} but for the Case-2 simulation ($\tau_1=10$ Myr, $\tau_2=30$ Myr, 1000 Plutos).}
\label{case2}
\end{figure}

Figures \ref{case1} and \ref{case2} show the orbital distribution of distant KBOs obtained in the Case-1 and Case-2 
simulations. The focus is on the orbits between 50 and 100 au. The first thing to be noted in these figures is that the
distribution of orbits with $q>40$ au has a very specific structure with concentrations near Neptune's mean motion resonances 
(MMRs), specifically the 3:1, 4:1, 5:1 and 6:1 MMRs. Additional concentrations are seen near the 7:2, 9:2, 11:2 and weaker 
resonances. In Case 1, the orbits fill a semimajor axis interval that starts some 2 AU on inside of the present resonant 
locations (except for 6:1 MMR where the orbits are more concentrated). In Case 2, the semimajor axis distributions
are more tightly concentrated near resonances. These findings are consistent with the results of Kaib \& Sheppard (2016). 

\begin{figure}
\epsscale{0.6}
\plotone{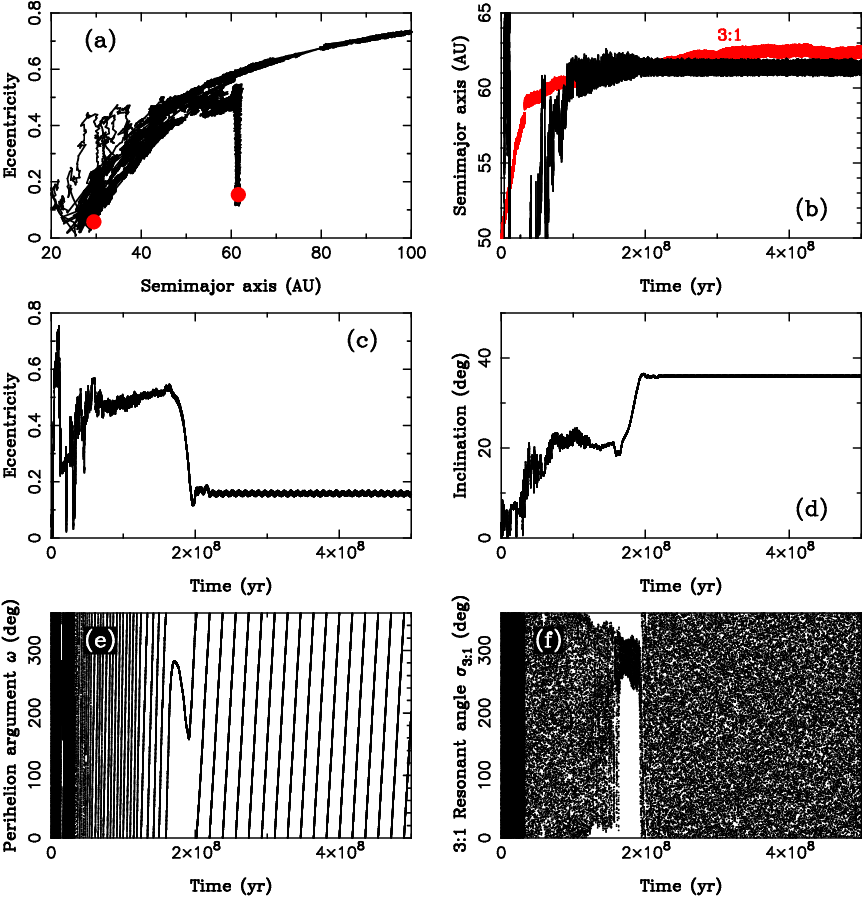}
\caption{An example of orbital evolution that ended with a detached orbit near the 3:1 MMR with Neptune. The red dots
in panel (a) show the initial ($a=29.47$ au, $e=0.058$, $i=3.4^\circ$) and final orbit ($a=61.24$ au, $e=0.160$, 
$i=36.0^\circ$). During the first stage of integration, the body is scattered by Neptune to an orbit with $a>50$ au, 
$i\simeq20^\circ$ (panel d) and large eccentricity (panel c). It subsequently becomes trapped in the 3:1 MMR
with Neptune, shown by the red line in panel b. The libration of the resonant angle $\sigma_{3:1}=3\lambda-\lambda_N-
2\varpi$, where $\lambda$ and $\lambda_N$ are body's and Neptune's mean longitudes, and $\varpi$ is the perihelion
longitude, occur between $t=180$ and 200 Myr (panel f). The resonant orbit is affected by Kozai cycles (panel e)
during which the eccentricity decreases and inclination increases, and the orbit decouples from Neptune. Finally, since
Neptune is migrating, the orbit drops from the 3:1 MMR and ends up $\simeq$1.4 au below the present resonance ($a=62.6$ AU).}
\label{example}
\end{figure}

The near-resonant orbits with $q>40$ au are created when the disk objects are scattered outward by Neptune and interact 
with resonances (Figure \ref{example}). The secular dynamics inside MMRs, mainly the Kozai cycles (Gomes 2003, Brasil et al. 2014), 
produce large oscillations of $e$ and $i$. When Neptune is still migrating, these resonant objects can be released from resonances 
with $q>40$ au and remain on stable orbits in the detached disk. The vast majority of these orbits are not inside the 
resonances today (the resonant angles do not librate).\footnote{Here we opt for not discussing the 5:2 resonance in 
detail, mainly because there is still some
disagreement about how large the population of objects inside the 5:2 resonance actually is (e.g., Volk et al. 2016, 
Sheppard et al. 2016). A large number of objects end up near the 5:2 resonance in our simulations (Figures \ref{case1} 
and \ref{case2}). A careful analysis shows that only a fraction of these objects are {\it inside} the 5:2 resonance today
($\simeq$50 particles in both the Case-1 and Case-2 simulations show sustained 5:2 resonant librations in an extended 
10-Myr simulation). This indicates the 5:2 implantation efficiency  $\simeq 5\times10^{-5}$, roughly 1/4 of the 3:2 
implantation efficiency in Case 1 with 4000 Plutos (NV16).}  

The resonant fingers shown in Figures \ref{case1} and \ref{case2} are a specific prediction of a model with the slow migration of 
Neptune (Nesvorn\'y 2015a). The fast migration ($\tau<10$ Myr) does not produce these fingers because there is not enough time with the 
fast migration for the secular cycles to act to raise the perihelion distance. Then, when Neptune stops migrating, all captures 
in resonances become temporary and the perihelion distances do not drop below 40 AU. Furthermore, the high-eccentricity phase 
of Neptune, investigated by Levison et al. (2008), would produce a different structure of the detached disk with $q>40$ au, where 
there is no strong preference for the resonant orbits.

\begin{figure}
\epsscale{0.6}
\plotone{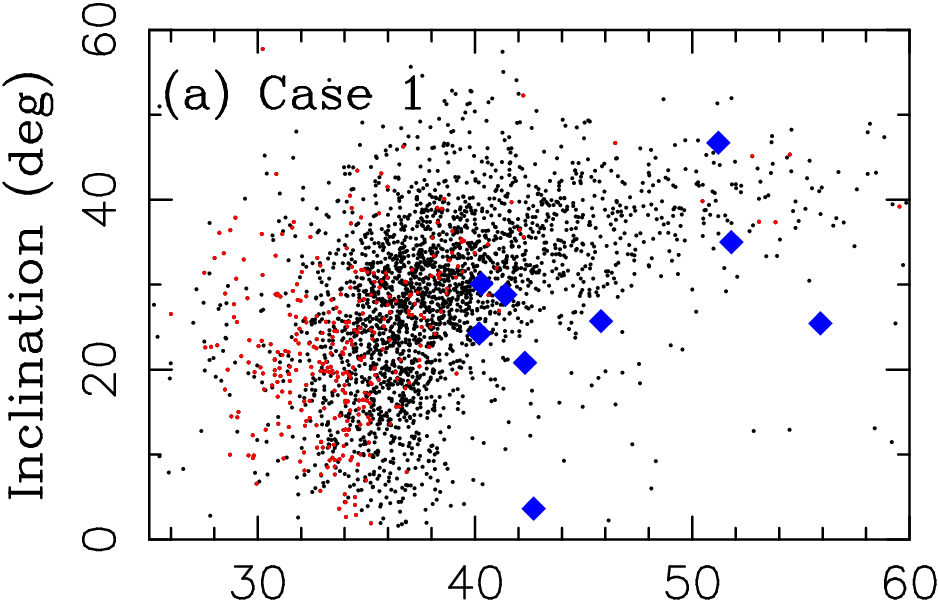}
\vspace{2.mm}
\plotone{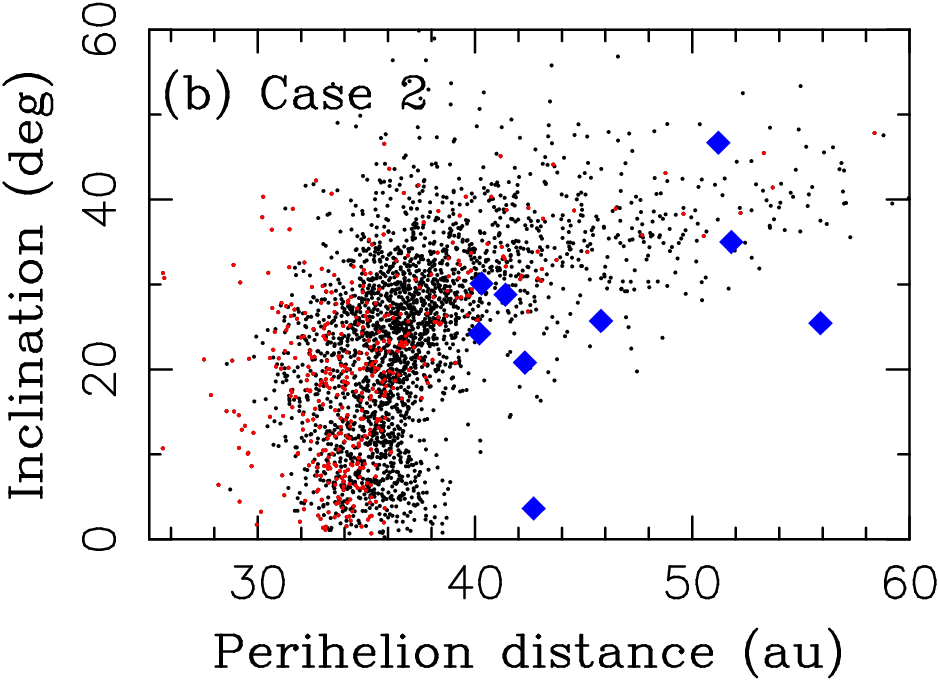}
\caption{The inclination distribution of distant KBOs ($a>50$ au) obtained in the Case 1 (a) and Case 2 (b) simulations. While 
the model objects with $q<40$ au have a wide range of inclinations, the ones with $q>40$ au generally have $i>20^\circ$. This 
corresponds pretty well to the orbital inclinations of KBOs detected by Sheppard et al. (2016) (here shown by blue diamonds).
The KBO with $q=42.7$ au and $i=3.6^\circ$, 2012 FH84, was probably not produced by the mechanism discussed here. Instead,
it may trace a continuation of Cold Classicals beyond 50 au (Sheppard et al. 2016). The scattering orbits are denoted by red 
dots.}
\label{incl}
\end{figure}

The recent detection of several new KBOs with $50<a<100$ au and $q>40$ au (Sheppard et al. 2016) are in line with our model 
predictions. These objects tend to concentrate toward resonances. There is 2015 FJ345, 2013 FQ28 and 2015 KH162 at 
the 3:1 resonance, 2014 FZ71 and 2005 TB190 at the 4:1 resonance and 2008 ST291 at the 6:1 resonance (Sheppard et al. 2016). 
In addition, all these objects, except of 2012 FH84, have high orbital inclinations ($i>20^\circ$), as expected if the Kozai 
cycles played role in their origin. We find from our simulations that the orbits with $q>40$ au indeed have large inclinations
(characteristically $\simeq$25-45 deg, with a clear correlation between $q$ and $i$; Figure \ref{incl}). This provides 
additional support for our model. The mean inclination of orbits with $q<40$ au is 25.7$^\circ$ in Case 1 and 21.3$^\circ$ 
in Case 2. The mean inclination of orbits with with $q>40$ au is similar ($\simeq35^\circ$) in both cases. The slower and 
grainier migration in Case 1 produced several low-inclination orbits ($i\lesssim10^\circ$) with $q>40$ au, while these 
orbits are almost non-existent in Case 2.    

The orbits of distant KBOs with $q>40$ au are not known well enough to establish whether they are resonant (which would contradict
predictions of our model) or non-resonant (which would support our model). Future observations will help to resolve this 
issue. In addition, the semimajor axis distributions of objects with $q>40$ au are sensitive to Neptune's migration speed
with faster migrations speeds implying more concentrated populations. This can be used, when the distributions are well characterized 
by observations, as a diagnostic of Neptune's migration speed (Kaib \& Sheppard 2016). Unlike the inclination distribution
considered in Nesvorn\'y (2015a), which can be used to mainly constrain the early stages of Neptune's migration, the semimajor 
axis distributions considered here should be more sensitive to the migration speed (and graininess) during the last $\sim$1 au of
Neptune's migration. If the independent arguments derived from Saturn's obliquity are valid (e.g., Vokrouhlick\'y \& Nesvorn\'y 
2015), Neptune's migration was very slow during the late stages ($\tau \sim 150$ Myr), thus favoring Case 1 over Case 2, and 
the semimajor axis distributions that are more spread on the inner side of resonances.   

In the nomenclature of Gladman et al. (2008), the SDOs can be divided into {\it scattering} objects (the ones that are currently 
scattering actively off Neptune; e.g., (15874) 1996 TL$_{66}$, Luu et al. 1997), and {\it detached} objects as being non-scattering 
SDOs with large eccentricity (e.g., (148209) 2000 CR$_{105}$, Buie et al. 2000). The scattering objects, defined as those whose 
semimajor axis changed more than 1.5 au in 10 Myr (Gladman et al. 2008), are denoted by red dots in Figures \ref{case1}
and \ref{case2}. We find that the slow migration model ($\tau\gtrsim10$ Myr)   
implies that the detached population should represent the majority of SDOs. Specifically, the implantation efficiency 
as a detached object with $50<a<100$ au is $2.0\times10^{-3}$ in both Case-1 and Case-2 simulations (Table 1). The 
implantation efficiency as a scattering object is much smaller, $3.7\times10^{-4}$ in Case 1 and $4.6\times10^{-4}$ in Case 2. 
This shows that the detached population should be $\simeq$5 times larger than the scattering population. All estimates
reported here apply to the part of the scattered disk between 50 and 100 au.  

Nesvorn\'y et al. (2013) estimated, using their model of Jupiter Trojan capture and the current population 
of Trojans, that the original planetesimal disk should have contained $\sim2\times10^7$ bodies with diameters $D>100$ km
(this assumes $7\times10^{-7}$ Trojan capture efficiency and the fact that there are 15 Jupiter Trojans with $H<8.7$, which corresponds 
to $D>100$~km for a 6\% albedo). If so, the detached population with $50<a<100$ au should have $\sim$40,000 objects with $D>100$~km. 
The scattering population in the same semimajor axis range should be smaller ($\sim$8,000 objects with $D>100$~km).
A careful consideration of observation biases will be required to understand how well this corresponds to the reality. 
   
Sheppard et al. (2016) estimated that there are $2400^{+1500}_{-1000}$ and $1600^{+2000}_{-1200}$ objects with $q>40$ au and
$D>100$ km at the 3:1 and 4:1 resonances. From our simulations, assuming $2\times10^7$ $D>100$ km objects in the original disk
and the implantation efficiencies reported in Table 1, we compute that there should be between $\sim$1600 (for Case 2) and 
$\sim$2400 (for Case 1) objects with $D>100$ km at the 3:1 resonance, and between $\sim$1000 (for Case 2) and 
$\sim$1400 (for Case 1) objects with $D>100$ km at the 4:1 resonance. This is consistent with the findings of Sheppard et al.
(2016). The populations are larger with slower migration (Case 1) because this case allows more time for the implantation of 
bodies into the detached disk. This dependence could, in principle, be used to constrain the migration speed of Neptune.
For that, however, we would need to consider a larger suite of integrations and have better 
observational constraints. According to our model, somewhat smaller populations should exists near the 5:1 and 6:1 resonances 
($\sim$500-1000 with $D>100$ km and $q>40$ au), and this trend should continue to weaker resonances beyond 100 au.  
\section{Conclusions} 
Our simulations with slow migration of Neptune (as required from the inclination constraint; Nesvorn\'y 2015a) lead to 
the formation of a prominent detached disk with substantial populations of objects concentrated at various MMRs 
with Neptune. This is an important prediction of the model, which is testable by observations. The current 
surveys are only starting to have a sufficient sensitivity to probe the orbital distribution of bodies with large 
perihelion distances (e.g., Shankman et al. 2016). 

Sheppard et al. (2016) reported several new objects in the detached disk between 50 and 100 au. They found 
that these objects are near Neptune's MMRs and have significant inclinations ($i>20^\circ$).
Interestingly, these findings are consistent with the predictions of our model with slow migration of Neptune.
The population census of near-resonant SDOs inferred from observations is also consistent with the model.

Our results imply that the detached population at 50-100 au should be $\simeq$5 times larger than the scattering population 
in the same semimajor axis range, which may have important implications for the origin of Jupiter-family comets. 
In addition, there seems to be a large population of objects with $q\simeq35$-40 au in the 5:1 MMR (Pike et al. 2015), 
which cannot be easily explained by the resonant sticking of scattering objects (Yu et al. 2015). Instead, we find it 
possible that these objects are the low-$q$, easier-to-detect part of the resonant populations that continue 
to $q>40$ au. 

\acknowledgements
The work of D.N. was supported by NASA's Emerging Worlds program and Brazil's Science without Borders program. 
D.V. was supported by the Czech Grant Agency (grant GA13-01308S). F.R. was supported by Brazil's Council of 
Research. All CPU-expensive simulations in this work were performed on NASA's Pleiades Supercomputer. We thank
Nathan Kaib and Scott Sheppard for discussions, and for making it possible to submit our article simultaneously 
with theirs (Kaib \& Sheppard 2016). We also thank an anonymous reviewer for useful suggestions.

\clearpage
\begin{table}
\centering
{
\begin{tabular}{lrr}
\hline \hline
              & Case 1           & Case 2   \\  
              & ($\times 10^{-4}$)         & ($\times10^{-4}$) \\
\hline
Detached      & 20               &  20  \\ 
Scattering    & 3.7              &  4.6 \\   
3:1           & 1.1              &  0.78 \\
4:1           & 0.70             &  0.47 \\
5:1           & 0.48             &  0.24 \\
6:1           & 0.32             &  0.26 \\                            
\hline \hline
\end{tabular}
}
\caption{The implantation probabilities in various target regions. These estimates were obtained by determining 
the number of particles that ended in target regions at 4.5 Gyr and dividing it by the number of particles 
in the original disk ($10^6$). The scattering objects are defined as those with the semimajor axis change 
$\Delta a > 1.5$~au in a 10 Myr integration. The detached objects have more stable orbits ($\Delta a < 1.5$~au).
The estimates are given for the populations with $50<a<100$ au. The resonant populations include objects both 
inside and close to the resonances with the latter ones being overhelmingly more common. The identification of these objects
in the distributions shown in Figures \ref{case1} and \ref{case2} was straightforward (we used appropriate 
semimajor axis ranges).}
\end{table}

\clearpage

\clearpage

\clearpage

\clearpage


\begin{thebibliography}

\bibitem[Batygin et al.(2012)]{2012ApJ...744L...3B} Batygin, K., Brown, M.~E., \& Betts, H.\ 2012, \apjl, 744, L3 

\bibitem[Batygin et al.(2011)]{2011ApJ...738...13B} Batygin, K., Brown, 
M.~E., \& Fraser, W.~C.\ 2011, \apj, 738, 13 

\bibitem[Batygin \& Brown(2016)]{2016AJ....151...22B} Batygin, K., \& Brown, M.~E.\ 2016, \aj, 151, 22 

\bibitem[Brasil et al.(2014)]{2014A&A...564A..44B} Brasil, P.~I.~O., Gomes, R.~S., \& Soares, J.~S.\ 2014, \aap, 564, A44 

\bibitem[Gladman et al.(2008)]{2008ssbn.book...43G} Gladman, B., Marsden, B.~G., \& Vanlaerhoven, C.\ 2008, The Solar System Beyond Neptune, 43 

\bibitem[Gomes(2003)]{2003Icar..161..404G} Gomes, R.~S.\ 2003, Icarus, 161, 404 

\bibitem[Gomes et al.(2004)]{2004Icar..170..492G} Gomes, R.~S., Morbidelli, 
A., \& Levison, H.~F.\ 2004, Icarus, 170, 492 

\bibitem[Kaib \& Sheppard(2016)]{2016arXiv160701777K} Kaib, N.~A., \& Sheppard, S.~S.\ 2016, arXiv:1607.01777 

\bibitem[Lawler et al.(2016)]{2016arXiv160506575L} Lawler, S.~M., Shankman, C., Kaib, N., et al.\ 2016, arXiv:1605.06575 

\bibitem[Levison 
\& Duncan(1994)]{1994Icar..108...18L} Levison, H.~F., \& Duncan, M.~J.\ 1994, Icarus, 108, 18 

\bibitem[Levison et al.(2008)]{2008Icar..196..258L} Levison, H.~F., 
Morbidelli, A., Vanlaerhoven, C., Gomes, R., 
\& Tsiganis, K.\ 2008, Icarus, 196, 258 

\bibitem[Luu et al.(1997)]{1997Natur.387..573L} Luu, J., Marsden, B.~G., Jewitt, D., et al.\ 1997, \nat, 387, 573 

\bibitem[Nesvorn{\'y}(2015)]{2015AJ....150...73N} Nesvorn{\'y}, D.\ 2015a, 
\aj, 150, 73 

\bibitem[Nesvorn{\'y}(2015)]{2015AJ....150...68N} Nesvorn{\'y}, D.\ 2015b, 
\aj, 150, 68 

\bibitem[Nesvorn{\'y} 
\& Morbidelli(2012)]{2012AJ....144..117N} Nesvorn{\'y}, D., \& Morbidelli, A.\ 2012 (NM12), \aj, 144, 117 

\bibitem[Nesvorny \& Vokrouhlicky(2016)]{2016arXiv160206988N} Nesvorn\'y, D., \& Vokrouhlicky, D.\ 2016 (NV16), arXiv:1602.06988 

\bibitem[Nesvorn{\'y} et al.(2013)]{2013ApJ...768...45N} Nesvorn{\'y}, D., Vokrouhlick{\'y}, D., \& Morbidelli, A.\ 2013, \apj, 768, 45 

\bibitem[Petit et al.(2011)]{2011AJ....142..131P} Petit, J.-M., Kavelaars, 
J.~J., Gladman, B.~J., et al.\ 2011, \aj, 142, 131 

\bibitem[Pike et al.(2015)]{2015AJ....149..202P} Pike, R.~E., Kavelaars, J.~J., Petit, J.~M., et al.\ 2015, \aj, 149, 202 

\bibitem[Shankman et al.(2016)]{2016AJ....151...31S} Shankman, C., Kavelaars, J., Gladman, B.~J., et al.\ 2016, \aj, 151, 31 

\bibitem[Sheppard et al.(2016)]{2016ApJ...825L..13S} Sheppard, S.~S., Trujillo, C., \& Tholen, D.~J.\ 2016, \apjl, 825, L13 

\bibitem[Trujillo \& Sheppard(2014)]{2014Natur.507..471T} Trujillo, C.~A., \& Sheppard, S.~S.\ 2014, \nat, 507, 471 

\bibitem[Vokrouhlick{\'y} \& Nesvorn{\'y}(2015)]{2015ApJ...806..143V} Vokrouhlick{\'y}, D., \& Nesvorn{\'y}, D.\ 2015, \apj, 806, 143 

\bibitem[Volk et al.(2016)]{2016AJ....152...23V} Volk, K., Murray-Clay, R., Gladman, B., et al.\ 2016, \aj, 152, 23 

\bibitem[Yu et al.(2015)]{2015DPS....4721108Y} Yu, T.~Y.~M., Murray-Clay, R., \& Volk, K.\ 2015, AAS/Division for 
Planetary Sciences Meeting Abstracts, 47, 211.08 

\end{thebibliography}
\end{document}